\documentclass[12pt,preprint]{aastex}

\shorttitle{NN2}
\shortauthors{Barris et al.}

\begin{document}
\title{The NN2 Flux Difference Method for Constructing
Variable Object Light Curves}

\author{Brian J. Barris\altaffilmark{1}, John L. Tonry\altaffilmark{1}, Megan C. Novicki\altaffilmark{1}, and W. Michael Wood-Vasey\altaffilmark{2}}

\altaffiltext{1}{Institute for Astronomy, University of Hawaii,
2680 Woodlawn Drive, Honolulu, HI 96822; 
{barris@ifa.hawaii.edu}, {jt@ifa.hawaii.edu}, {mnovicki@ifa.hawaii.edu}}

\altaffiltext{2}{Harvard-Smithsonian Center for Astrophysics,
60 Garden Street, Cambridge, MA 02138;
{wmwood-vasey@cfa.harvard.edu}}

\begin{abstract}
We present a new method for optimally extracting point-source time
variability information from a series of images.
Differential photometry is generally best accomplished by subtracting
two images separated in time, since this removes all constant objects
in the field.
By removing background sources such as the host galaxies of supernovae, 
such subtractions make possible the measurement of the proper
flux of point-source objects superimposed on extended sources.
In traditional difference photometry, a single image is
designated as the ``template'' image and subtracted from all other
observations.  This procedure does not take all the available
information into account and for sub-optimal template images may
produce poor results.  Given $N$ total observations of an object, we
show how to obtain an estimate of the vector of fluxes from the
individual images using the antisymmetric matrix of flux differences
formed from the $N(N-1)/2$ distinct possible subtractions and provide
a prescription for estimating the associated uncertainties.  We then demonstrate
how this method improves results over the standard procedure of
designating one image as a ``template'' and differencing against only
that image.

\end{abstract}


\section{Introduction}

The astronomical time domain provides unique insight into a range of
astrophysical phenomena.  Studies of variable stars yield information
about stellar structure and evolution as well as help to set the
extra-galactic distance scale.  Active Galactic Nuclei (AGN) reveal the
high-energy phenomena associated with the super-massive black holes
that reside at the centers of most galaxies.  Supernovae (SNe) and
Gamma-Ray Bursts (GRBs) provide a glimpse of the fantastic energies
released during the violent death throes of several types of stars.
Type~Ia supernovae (SNe~Ia) are of particular interest because their
use as ``standard candles'' has revealed the acceleration of the
expansion of the universe from an inferred cosmological constant-like
force~\citep{riess98,perlmutter99}.

Studies of variable sources require specific analysis methods that
are not necessary for non-variable sources.  Since it is often
difficult to detect variation in an object by simply inspecting
images, the standard procedure is to subtract images taken at
different times to remove objects with constant flux.
Photometrically variable objects are then obvious.  For the case of
SNe, one typically obtains a pair of observations separated in time to
allow for SNe not present in the first image to reach observable
brightness in the second (see, e.g., \citealt{perlmutter95} and \citealt{schmidt98}
 for a description of the method).  After
detection, additional observations are made to obtain the complete light
curve necessary for cosmological analysis \citep[see][]{phillips93,riess96}.  
In order to construct the light curve, it
is necessary for at least one observation (the ``template'' image) to contain
no SN flux.  Often this is the initial image used during discovery of
the SN.  In many instances, however, the SN is present at a faint
level in this image, so an additional observation, taken after the SN
has faded from view, is required.  The light curve is then calculated
by measuring the flux levels in subtractions of each image from the
designated template using, for example, the subtraction
procedure described by
\citet{alard98}.  This, the ``single-template method,'' is the
typical means of constructing light curves of SNe and other variable
sources.

However, this method has certain drawbacks.
The primary flaw is that the quality of any subtraction depends
greatly upon the two images involved.  If the template is of a poor
quality caused, for instance, by poor seeing or a low signal-to-noise
ratio (S/N), then $every$ subtraction will be degraded, with a
corresponding increase in the measured flux uncertainty, even if all
other images are of high quality.  Any flaw in the template creates a
systematic error for the entire light curve that is not detectable
from internal consistency checks or through comparison with another SN
light curve.

In order to alleviate this problem, we have developed a new method for
constructing light curves of photometrically variable objects.  Given
$N$ observations there are a total of $N(N-1)/2$ pairs of images that
can be subtracted together, only $N-1$ of which are performed in the
single-template method.  A matrix of flux differences can be
constructed from these subtractions and used to determine the flux at
each individual epoch.  This process removes the dependence on any
single observation, because all observations are treated equally as a
``template.''  We refer to this method as the ``N(N-1)/2'' method
(hereafter abbreviated NN2; see \citealt{novicki00} for an initial
description).

Section~\ref{sec-math} describes the mathematical underpinnings
of NN2.  In Section~\ref{sec-example} we demonstrate the efficacy of
the method using simulated SNe inserted into images used during an
actual high-redshift SN survey.  Section~\ref{sec-conclusions}
gives our conclusions.

\section{Mathematical Basis of the NN2 Method}
\label{sec-math}

We assume that we start with $N$ observations of an object, so that
one may construct from all pairs of subtractions an $N\times N$
antisymmetric matrix $A$ of flux differences that we wish to analyze
as a ``vector-term difference.'' 
In other words, we want to find an $N$-vector $V$ of fluxes such that
\begin{equation}
A_{ij} = V_j - V_i.
\end{equation}
We also assume that we have a symmetric $N\times N$ error matrix $E$
that expresses our uncertainty in each term of $A$.  As we shall see,
this matrix may not be easy to generate, and its interpretation may be
somewhat ambiguous.  However, one can imagine generating an error
matrix by the following procedure.

In each of the difference images, we measure a flux for the object in question.  
In general this measurement consists of 
fitting a fixed point-spread function (PSF) profile at the location
of the object by adjusting the amplitude of the PSF (both positive or
negative) and the local background level.  
The PSF profile may be
obtained from a suitable star in the original image while the location
of the variable source may be determined by summing all the difference images (adjusted
to keep the sign of the object positive) and fitting the location in
this sum.  Once we have a flux measurement, we can insert copies of
the object at nearby empty regions of the difference image and repeat
the procedure.  The mean of the recovered fluxes indicates whether
there is a bias in the measurement, and the scatter may be used as a
term $E_{ij}$ in the error matrix.

The crux of the NN2 method is the distillation of the photometric
measurements from the full set of $N(N-1)/2$ subtractions to
a lightcurve, $V$, that represents our best understanding of the behavior
of the object under consideration.  As long as it is consistent,
the exact procedure for measuring the flux on the difference images
is not central to the NN2 method we present here.

In order to find an optimal $V$, we wish to minimize the quantity
\begin{equation}
\chi^2 = \sum_{i,j;i<j} {(-A_{ij} + V_j - V_i)^2 \over E_{ij}^2}
\label{eq:chi2}
\end{equation}
This construction may not be entirely appropriate depending on the
errors in the flux differences $A_{ij}$.  Ideally, if we possessed an
extremely high-quality template with no SN flux present and applied an
optimal subtraction procedure, the errors would be primarily due to
photon counting statistics (see \citealt{alard98} for a
discussion).  We would expect these errors to be uncorrelated and
would simply wish to employ the single-template method to construct
the SN light curve.  As mentioned in the introduction, in practice
there are nearly always imperfections associated with the template
image that remove us from this idealized regime.  These template
errors introduce correlations in the individual flux measurement
errors that are difficult to quantify and are typically assumed to be
negligible in SN light-curve analysis.  The use of the NN2 method,
however, will introduce further correlations as a result of the common
images in the various subtractions (for instance, the error in
$V_1-V_2$ will be anti-correlated with the error in $V_2-V_3$ due to
the
common error in $V_2$).  Although we believe that the use of the NN2
method will improve the ability to accurately recover variable object
light curves, one should recognize that the NN2 method is expected to introduce
these additional correlations to the fluxes measured from the various
subtraction images, and so the $\chi^2$ given above is not technically
appropriate.  Errors due to systematics in the subtraction procedure,
such as those associated with template or software imperfections,
would be expected to be effectively uncorrelated, and if they were
dominant then Eq.~\ref{eq:chi2} would indeed represent the proper $\chi^2$.  
With these caveats in mind, we will proceed to use the definition of
$\chi^2$ as given in Eq.~\ref{eq:chi2} as the basis of the NN2 method.
Tests of its ability to recover accurate light-curve information in
the following section demonstrate its effectiveness in practice.

However, we need to make one minor modification to our
$\chi^2$ because the $\chi^2$ defined in Eq.~\ref{eq:chi2}
is degenerate to the addition of a constant to the
$V$ vector---geometrically, $\chi^2$ is constant along the line $\sum
\hat i$.  In order to lift this degeneracy and permit us to solve for
$V$, we add a term to $\chi^2$ that is quadratic in the degenerate
direction, so that
\begin{equation}
\chi^2 = \sum_{i,j;i<j} {(-A_{ij} + V_j - V_i)^2 \over E_{ij}^2} + 
{(\sum_i V_i)^2 \over \langle E\rangle^2},
\label{eq:final_chi2}
\end{equation}
where $\langle E\rangle$ is a suitable typical uncertainty; for
example, 
\begin{equation}
{1\over \langle E\rangle^2} = {2\over N(N-1)}\,\sum_{i,j;i<j} {1\over E_{ij}^2}.
\end{equation}
Our solution will therefore have $\sum_i V_i = 0$.  
This construction explicitly forces one to determine an accurate 
zero flux level at a later stage.
In the single-template method this zero flux level is generally 
implicitly determined by assuming that the object of
interest has zero flux in the template image.
This same assumption can similarly be used in the NN2 method,
but more sophisticated methods involving comparisons of
many different images can also be invoked.
If the absolute brightness of the variation being studied is
important, the NN2 method clearly cannot free one from the
requirement of having a fiducial image to measure the zero flux level.
This is a fundamental limitation of any differential photometry method
as the information is simply not available without such a fiducial
image.
However, even in the absence of a fiducial image, the NN2 method will produce
a sensible and accurate relative lightcurve.

We now seek to solve for our lightcurve vector $V$ by minimizing $\chi^2$ with respect to $V$:
\begin{eqnarray}
0 & = & {\partial \chi^2 \over \partial V_k} \\
  & = & 2\sum_{i,j;i<j} {(-A_{ij} + V_j - V_i) \over E_{ij}^2} (\delta_{jk} - \delta_{ik}) + 2\sum_i {V_i\over \langle E \rangle^2}.
\label{eq:chi2v}
\end{eqnarray}
Exploiting the antisymmetry of $A$ and the symmetry of $E$ we can rewrite Eq.~\ref{eq:chi2v} as
\begin{equation}
0 = 2\sum_{i;i\neq k} {(-A_{ik} + V_k - V_i) \over E_{ik}^2}
    + 2\sum_i {V_i \over \langle E \rangle^2}.
\end{equation}
These $N$ equations can be solved for $V$ by inverting a matrix $C$:
\begin{equation}
  \sum_{i;i\neq k} {A_{ik} \over E_{ik}^2} = \sum_i C_{ik} V_i
\end{equation}
where
\begin{equation}
  C_{ik} = {-1 \over E_{ik}^2} + \sum_j{1 \over E_{kj}^2} \delta_{ik} +
           {1 \over \langle E \rangle^2}.
\end{equation}

The inverse of this curvature (Hessian) matrix $C$ yields uncertainties
for $V$ from the square root of the diagonal elements as well as covariances
from normalizing the off-diagonal elements by the two diagonal terms
(under the assumption that the error matrix truly does represent
Gaussian, independent uncertainties for each of the terms of the
antisymmetric difference matrix).

An alternative approach to calculating uncertainties in $V$ stems from
assuming that there is a vector $\sigma$ such that
\begin{equation}
E_{ij}^2 = \sigma_i^2 + \sigma_j^2.
\end{equation}
Under this assumption, we seek to minimize
\begin{equation}
\chi_e^2 = \sum_{i,j;i<j} \left(-E_{ij}^2 + \sigma_i^2 + \sigma_j^2 \right)^2.
\end{equation}
The minimization condition is
\begin{eqnarray}
0 & = & {\partial \chi_e^2 \over \partial \sigma_k^2 } \\
  & = & 2\sum_{i,j;i<j} \left(-E_{ij}^2 + \sigma_i^2 + \sigma_j^2 \right)\left(\delta_{ik}+\delta_{jk}\right) \\
  & = & 2\sum_{i;i\neq k} \left(-E_{ik}^2 + \sigma_i^2 + \sigma_k^2 \right).
\end{eqnarray}
These $N$ equations are solved by inverting a matrix $D$
\begin{equation}
  \sum_{i;i\neq k} E_{ik}^2 = \sum_i D_{ik} \sigma_i^2
\end{equation}
where
\begin{equation}
  D_{ik} = 1 + (N-2) \delta_{ik}.
\end{equation}

After solving for $V$ and $\sigma$, we can evaluate the quality
of the fit by comparing $\chi^2$ to the number of degrees of freedom,
\begin{equation}
N_{\mathrm dof} = {N(N-1)\over2} - (N-1).
\end{equation}
This $N_{\mathrm dof}$ comes from the number of data points,
$N(N-1)/2$, minus the number of model parameters, $N-1$.  Recall that
we've explicitly required $\sum_i V_i = 0$, so that the number of model
parameters is $N-1$ rather than $N$.

Having outlined the basic method,
we now discuss a fundamental uncertainty in the NN2 process.
We can imagine two types of error that will cause $V$ to differ from
the true flux values.  The first, which we term ``external error,'' is
intrinsic to the images themselves.  For example, if the object has a
positive statistical fluctuation in flux in one image or is corrupted
by a cosmic ray that happens to coincide with the position of the
object on the detector, this error will propagate
through the entire differencing and analysis procedure.  It is
possible to obtain an antisymmetric difference matrix that is an
exact vector-term difference ($\chi^2 = 0$), but the solution vector
will still contain errors.  The second type of error, which we call
``internal error,'' is caused by the procedure of generating the
antisymmetric matrix.  One might imagine a set of images that have no flux
error whatsoever, but through errors in convolving, differencing, or
flux fitting, an antisymmetric matrix may be created that is not a perfect
vector-term difference and for which $\chi^2 > 0$.

Roughly speaking, one might expect that if the error matrix $E$
consists entirely of external errors the resulting $\sigma$ terms will
all be approximately $E/\sqrt{2}$, since $E$ is the quadrature sum of
two $\sigma$ terms.  Alternatively, if the error matrix is purely
internal error the $\sigma$ terms might be expected to be
approximately $E/\sqrt{N}$, since each term in $V$ comes from
comparison with $N-1$ other images.  In the case of external errors, the
uncertainties derived from the $\chi_e^2$ analysis are correct.  In
the internal error case the uncertainties obtained from the covariance
matrix derived from the $\chi^2$ analysis are likewise appropriate.

It is not clear how to disentangle these different sorts of errors.
The procedure suggested above of dropping copies of the object into
each difference image and evaluating the scatter of the result will be
sensitive to each sort of error, but it is possible to imagine cases
where this procedure is unsatisfactory.  We suggest that the errors
provided in the $E$ matrix be interpreted as external errors and taken
seriously as such.  Thus, the vector $V$ is assigned an external
uncertainty equal to the $\sigma$ vector.  However, in order to handle
a situation where $\chi^2 / N_{\mathrm dof}$ is much greater than 1
(i.e., where the antisymmetric matrix is simply {\it not} well represented as
a vector-term difference), we suggest also creating an internal
uncertainty vector $\tau$ that is obtained from the diagonal terms of
the covariance matrix, scaled by $\chi^2 / N_{\mathrm dof}$:

\begin{equation}
  \tau_k = \left( C^{-1}_{kk} \; {\chi^2 \over N_{\mathrm dof}} \right)^{1/2}
\end{equation}
The total uncertainty is then the quadrature sum of $\sigma$ and
$\tau$.  Note that this approach implicitly assumes that the internal and
external errors are uncorrelated and are proportional to one another 
as well as the provided $E$ matrix.

For problems where $\chi^2/N_{\mathrm dof}$ is near unity without adjustment,
the $\tau$ vector will be smaller than the $\sigma$ vector by
approximately $\sqrt{2/N}$ and will make a fairly small contribution
to the total uncertainty.  When $\chi^2/N_{\mathrm dof} \ll 1$ (i.e., the
antisymmetric matrix is very closely represented by the vector-term
difference), the $\tau$ vector will be negligible.  However, when
$\chi^2/N_{\mathrm dof} \gg 1$, the $\tau$ vector will act to correct
$\chi^2/N_{\mathrm dof}$ to approximately unity, and this procedure will
provide reasonable uncertainties, even though $E$ may be much too
small.

\section{Demonstration of Improved Accuracy in Recovering Variable Object Light Curves}
\label{sec-example}

The first extensive use of the NN2 method we have developed here
occurred during the SN-search component of the IfA Deep Survey~\citep{barris04a}, although we also employed it to a limited extent in a
previous SN survey by \citet{tonry03}.  The IfA Deep Survey was
undertaken primarily with Suprime-Cam~\citep{miyazaki98} on the
Subaru 8.2-m telescope and was supplemented with the 12K camera~\citep{cuillandre99} on the Canada-France-Hawaii 3.6-m telescope.
Scores of high-redshift SN candidates were discovered~\citep{barris01,barris02} with 23 confirmed as SNe~Ia.  We here present several
tests we performed to demonstrate the improved performance of NN2
relative to the single-template method.

In order to make a controlled test of the effectiveness of NN2 vs. the
single-template method, we inserted artificial SNe into the survey
images.
The light curves of these objects
consisted of a linear ramp-up and ramp-down in brightness over the
time period covered by the survey observations.  The timing of the
light-curve maxima were selected at random and could lie within or
outside of the survey period.  The simulated SNe were laid down in a
regular grid across the survey area, and all pairs of images
were subtracted.  Object-detection software was then run on all subtraction
images to detect photometrically variable objects (both real objects
and the artificial SNe) and to construct the NN2 flux difference matrix.
The positions of both real and artificial SNe were fit by the
object-detection software as described in Section~\ref{sec-math}.
Since we knew the true light-curve properties used to
create the synthetic SNe, we could calculate the root-mean-square 
scatter (RMS) around this artificial light curve using both the NN2 flux calculation
and the single-template method with every individual observation as
the template (this latter is equivalent to taking the flux values from a
single column of the NN2 flux difference matrix).

We inserted approximately 2000 simulated SNe into the $I$-band
observations of each of four $\sim$0.5 square-degree fields from the
IfA Deep Survey, spanning a peak magnitude range of approximately
$m_I=21$--$25$.  We used a predefined grid of positions to insert the
simulated SNe, without taking into consideration the presence of
actual objects nearby that would cause problems for detection and
accurate photometric measurements.  The small fraction that were
so affected were accordingly not used in the final analysis.

\begin{figure}
\epsscale{1.0}
\plotone{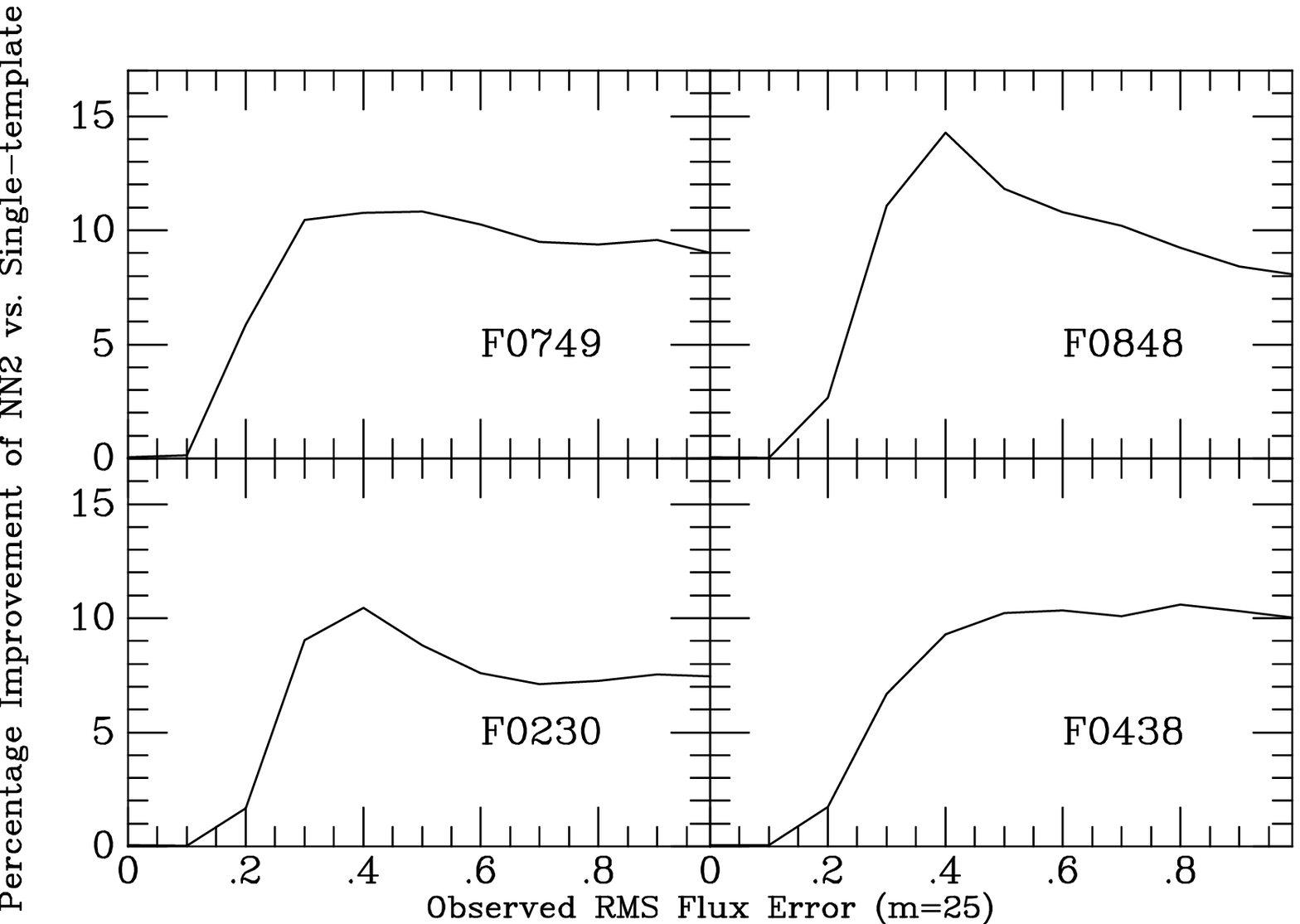}
\caption{
The improvement in the cumulative RMS distribution for the NN2 technique 
compared with that of the single-template method for each of four
survey fields is shown here as the extra fraction of NN2-derived
light curves, compared to the conventional light curves, that have an
observed RMS flux error less than the abscissa value.
For each field the NN2 cumulative fraction is larger at all
RMS values, demonstrating its improved performance in accurate 
light-curve recovery compared to the single-template method.}
\label{fig:fakehisto}
\end{figure}

Figure~\ref{fig:fakehisto} shows the percentage improvement in the cumulative distributions of the RMS (in
flux units) from the NN2 method over the set of all RMS values
from the single-template method from the four survey fields (RMS
values are calculated from flux measurements scaled so that a value
of $\mathrm{flux}=1$ corresponds to a magnitude of $25$).  The cumulative fraction for the
NN2 method is larger than that for the single-template method distribution
at all values of RMS, indicating that the NN2 method does indeed tend
to yield $smaller$ RMS values.  The NN2 method more accurately recovers the
actual light curve of these variable objects.

Since one could imagine that certain templates of very high
quality could outperform the NN2 method while the collection of all
single-template measurements, as shown in Fig.~\ref{fig:fakehisto}, does not, we next examine the
relationship between the NN2 RMS and the single-template RMS for individual
observations.  To illustrate this comparison, we will concentrate on only one of
the survey fields, although details of our investigation of the entire
survey area can be found in \citet{barris04b}.
Table~\ref{table:obsinfo-0438} contains
relevant information for the 16 observations of the selected field, f0438.
This field is representative of the entire survey area, though it is
notable that it contains an observation that was quite strongly
affected by clouds (Observation 4), as seen by its unusually bright
zero-point magnitude.  Also noteworthy is Observation 11, taken in
poor seeing conditions.  We would expect the performance of the
single-template method using these observations to be poor in
comparison to the NN2 procedure.

\begin{figure}
\epsscale{1.0}
\plotone{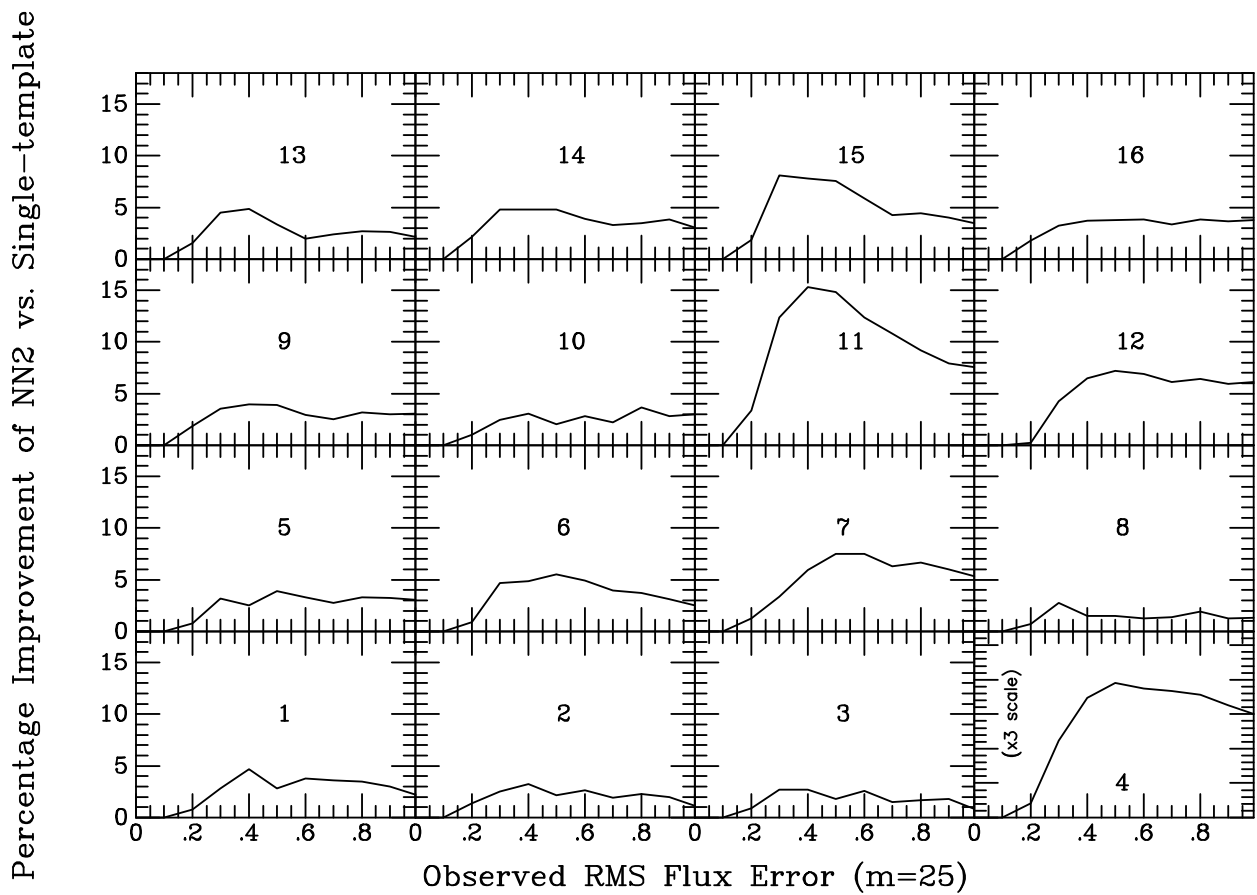}
\caption{
The percentage improvement in the cumulative RMS distribution for the NN2 technique 
compared with that of the single-template method 
for each of the 16 observations of a representative field, f0438, is shown here as in Fig.~\ref{fig:fakehisto}.  
Particularly noteworthy are the poor results of using as the single
template either observations 4 or 11, which suffered from heavy cloud
cover and poor seeing, respectively.
The vertical scale of the sub-plot for Observation 4 has been
accordingly scaled up by a factor of 3.
For templates of higher quality, the difference in relative
performance is less significant.  Nonetheless, NN2 outperforms the
single-template method for any choice of the single template image.
}
\label{fig:rmscumfake-0438}
\end{figure}

In Table~\ref{table:obsinfo-0438} we demonstrate that the typical RMS
obtained with the NN2 method for the set of 1775 simulated SNe is
smaller than the single-template method RMS using $every$ observation
of the selected field.  The improvement is generally fairly small,
ranging from $\sim5-10$\%.  We demonstrate below that these
differences are statistically significant.  The use of either Observations 4
and 11 as single templates, as expected, produces substantially worse
results relative to the NN2 method than the other observations.  For
these observations the improvement due to the NN2 method is
substantially larger than 10\%.  Figure~\ref{fig:rmscumfake-0438} shows
graphically the percentage difference in the cumulative RMS distributions, similar to 
Figure~\ref{fig:fakehisto}, for each individual observation of f0438.

Having demonstrated the improved performance of the NN2 method, we can
test the statistical significance of the differences between the NN2
RMS values and those calculated via the single-template method and
examine whether these differences indicate an actual difference in the
distributions of the results from the two methods.  To do so we use
the non-parametric Kolmogorov-Smirnov (K-S) test, with results given
in Table~\ref{table:ksinfo-0438}.  For the sample of all
single-template RMS values compared to NN2, the K-S probability value
is $\lesssim5\times10^{-9}$, indicating with strong confidence that the
distributions are different.  We also divide the sample into magnitude
bins, since the relative behavior of NN2 RMS to single-template RMS is
expected to be sensitive to the object's S/N and hence to the magnitude for 
a given sensitivity.

The K-S probability values for three approximately equal
magnitude bins show that for each of the subsamples the difference
between NN2 and the single-template method is statistically
significant, increasingly so at fainter magnitudes.  Finally, we
compare each observation individually with the NN2 method and see
again that the observed improvement in RMS with NN2 is highly
significant for nearly all observations (the only obvious potential
exception is observation 8).  These K-S probability values demonstrate
that the NN2 method is not distributed identically to the
single-template method, and the differences in median values given in
Table~\ref{table:obsinfo-0438} are indeed indicative of statistically
significant differences in the distributions.  
This test confirms that the NN2 method truly does produce improved results
in generating differential light curves.

\section{Conclusions}
\label{sec-conclusions}

We have described the mathematical foundation of a new method for
constructing the light curves of photometrically variable objects.
This method uses all $N(N-1)/2$ possible subtractions involving $N$
images in order to calculate a vector of fluxes of the variable
object and offers a powerful alternative to the single-template method 
that is in standard use for studying variable sources.

If one has a data set with a limited number of good fiducial
observations, the NN2 method will outperform any single-template
subtraction approach.
For cases where a large number of fiducial observations are available
to construct a deep template image, the NN2 method and the
single-template approach using this deep template should yield
comparable results.  In this situation we would encourage the use of 
both methods to provide additional checks and constraints on the
differential light curve.

We have tested the performance of the NN2 method by inserting
artificial SNe into images from the IfA Deep Survey and comparing the
RMS scatter from flux measurements using the two different methods.
We find that the RMS from the NN2 method is better than the
single-template RMS for the large majority (typically 65\%-72\%) of
the SNe for every possible template.  The median values for the ratio
of NN2 RMS (in flux units) to single-template RMS measurements are
typically $0.93$--$0.96$, demonstrating that the NN2 method results in
a $\sim5$\% improvement in the accuracy of the recovered light curve
for these observations.

Using Kolmogorov-Smirnov statistics, we have demonstrated that these
differences are significant, reflecting an actual difference between
the performance of the two methods.  We find extremely high
probabilities that the NN2 RMS is distributed significantly
differently from the single-template RMS values.  This difference and
improvement in RMS holds even
for the very high quality templates that would be considered ideal for
the single-template method.

We therefore make the following conclusions:

1. For the IfA Deep Survey observations, use of the NN2 method
   typically results in a 5-10\% improvement in the RMS of the recovered
   light curve in comparison to the single-template method.

2. For observations that have a large external error, such as those
   taken under poor conditions, the NN2 method results in a
   substantial improvement ($\gg10\%$) over the single-template method.

3. When working with high-quality observations, with small external
   error, the internal errors (such as those due to implementation of the
   subtraction process) dominate.  If these errors are large, 
   the NN2 method
   should outperform the single-template method to a large degree.  If
   these errors are kept small, as we believe is possible based our
   extensive experience with SN surveys, then the NN2 method will
   result in a modest but significant improvement in accuracy of 
   light-curve recovery.

In summary, the NN2 method we present here maximizes the time
variability information contained in a series of observations by
using the relative differences between all pairs of images to
construct the optimal differential light curve.
references

The source code for our implementation of the NN2 method presented
here is available at \url{http://www.ctio.noao.edu/essence/nn2/}.

\acknowledgments

This work was supported in part by grant AST-0443378 from the United
States National Science Foundation.

\clearpage

\clearpage

\clearpage

\begin{deluxetable}{cccccc}
\tablewidth{0pc}
\tablecaption{F0438 Observational Information and Light-Curve Recovery Comparison}
\tablehead{
\colhead{Obs. no.} & \colhead{MJD} &
\colhead{seeing\tablenotemark{1}} & 
\colhead{ZP\tablenotemark{2}} &
\colhead{Median RMS\tablenotemark{3}} &                                
\colhead{Median(RMS$_{NN2}/$RMS$_{temp.}$)\tablenotemark{4}} }
\startdata
  1  &  52164.57 &   0.75  &  30.48  &   0.441  &    0.951    \\  
  2  &  52176.49 &   0.72  &  30.05  &   0.432  &    0.963    \\  
  3  &  52191.50 &   0.71  &  30.12  &   0.427  &    0.970    \\  
  4  &  52198.56 &   0.59  &  27.63  &   0.810  &    0.625    \\  
  5  &  52204.59 &   0.63  &  30.46  &   0.424  &    0.973    \\  
  6  &  52225.39 &   0.62  &  30.17  &   0.445  &    0.960    \\  
  7  &  52231.38 &   0.51  &  30.46  &   0.455  &    0.944    \\  
  8  &  52232.35 &   0.68  &  30.21  &   0.419  &    0.970    \\  
  9  &  52236.34 &   0.94  &  30.36  &   0.440  &    0.938    \\  
 10  &  52252.35 &   0.83  &  30.59  &   0.429  &    0.957    \\  
 11  &  52263.38 &   1.14  &  30.33  &   0.515  &    0.839    \\  
 12  &  52283.39 &   0.68  &  30.51  &   0.460  &    0.946    \\  
 13  &  52288.23 &   0.84  &  30.65  &   0.438  &    0.941    \\  
 14  &  52289.39 &   0.93  &  30.27  &   0.448  &    0.936    \\  
 15  &  52323.35 &   0.97  &  30.23  &   0.468  &    0.908    \\  
 16  &  52369.25 &   0.58  &  29.58  &   0.432  &    0.943    \\  
\enddata
\tablenotetext{1} {Arcseconds}
\tablenotetext{2} {Magnitude zero-point of the processed images from
each observation}
\tablenotetext{3} {RMS calculated for flux units, with flux=1
corresponding to $m=25$.  Median RMS value for the NN2 method is 0.422.} 
\tablenotetext{4} {Median value of the ratio of the RMS as calculated
  with the single-template method to the RMS as calculated with the NN2 method.  A
  total of 1775 simulated SNe were inserted in the survey images for
  this field.}
\label{table:obsinfo-0438}
\end{deluxetable}

\begin{deluxetable}{rrccl}
\tablewidth{0pc}
\tablecaption{Field 0438 Kolmogorov-Smirnov Statistics}
\tablehead{
\colhead{N1} & \colhead{N2} & \colhead{cut criterion} & 
\colhead{K-S statistic} & \colhead{K-S probability\tablenotemark{1}}}
\startdata
   1775   &    28400 &   all                              &  0.0769   &   4.7430e-09   \\
    482   &     7712 &   $\ \ \ \ \ \ \ \ \ m < 23.0$     &  0.0809   &   4.9083e-03   \\
    617   &     9872 &   $23.0\leq m \leq24.5$            &  0.0897   &   1.5707e-04   \\
    676   &    10816 &   $24.5 < m \ \ \ \ \ \ \ \ \ $    &  0.1330   &   2.5150e-10   \\
   1775   &    1775  &   $\ $observation  1               &  0.0524   &   1.4698e-02   \\
   1775   &    1775  &   $\ $observation  2               &  0.0462   &   4.3869e-02   \\
   1775   &    1775  &   $\ $observation  3               &  0.0411   &   9.6888e-02   \\
   1775   &    1775  &   $\ $observation  4               &  0.3977   &   $<$1.0e-30    \\
   1775   &    1775  &   $\ $observation  5               &  0.0445   &   5.7717e-02   \\
   1775   &    1775  &   $\ $observation  6               &  0.0608   &   2.6510e-03   \\
   1775   &    1775  &   $\ $observation  7               &  0.0845   &   5.6244e-06   \\
   1775   &    1775  &   $\ $observation  8               &  0.0332   &   2.7611e-01   \\
   1775   &    1775  &   $\ $observation  9               &  0.0518   &   1.6326e-02   \\
   1775   &    1775  &   observation 10                   &  0.0417   &   8.9127e-02   \\
   1775   &    1775  &   observation 11                   &  0.1639   &   2.5672e-21   \\
   1775   &    1775  &   observation 12                   &  0.0794   &   2.4901e-05   \\
   1775   &    1775  &   observation 13                   &  0.0575   &   5.4234e-03   \\
   1775   &    1775  &   observation 14                   &  0.0710   &   2.4226e-04   \\
   1775   &    1775  &   observation 15                   &  0.0885   &   1.6592e-06   \\
   1775   &    1775  &   observation 16                   &  0.0541   &   1.0650e-02   \\
\enddata
\tablenotetext{1} {The probability that identically distributed
distributions will exhibit a value for the K-S statistic larger than
that observed. }
\label{table:ksinfo-0438}
\end{deluxetable}

\end{document}